

\def\oneandahalfspace{\baselineskip=16pt plus 1pt
\lineskip=2pt\lineskiplimit=1pt}

\def\nofirstpagenoten{\nopagenumbers\footline={\ifnum\pageno>1\tenrm
\hss\folio\hss\fi}}
\def\nofirstpagenotwelve{\nopagenumbers\footline={\ifnum\pageno>1\twelverm
\hss\folio\hss\fi}}
\def\leaderfill{\leaders\hbox to 1em{\hss.\hss}\hfill}


\parindent=20pt
\def\narrow{\advance\leftskip by 40pt \advance\rightskip by 40pt}

\def\nonarrower{\advance\leftskip by -40pt\advance\rightskip by -40pt}

\def\boxit#1{\vbox{\hrule\hbox{\vrule\kern3pt
        \vbox{\kern3pt#1\kern3pt}\kern3pt\vrule}\hrule}}

\def\gtorder{\mathrel{\raise.3ex\hbox{$>$}\mkern-14mu
             \lower0.6ex\hbox{$\sim$}}}
\def\ltorder{\mathrel{\raise.3ex\hbox{$<$}|mkern-14mu
             \lower0.6ex\hbox{\sim$}}}
\def\dalemb#1#2{{\vbox{\hrule height .#2pt
        \hbox{\vrule width.#2pt height#1pt \kern#1pt
                \vrule width.#2pt}
        \hrule height.#2pt}}}

\font\twelvett=cmtt12 \font\twelvebf=cmbx12
\font\twelverm=cmr12 \font\twelvei=cmmi12 \font\twelvess=cmss12
\font\twelvesy=cmsy10 scaled \magstep1 \font\twelvesl=cmsl12
\font\twelveex=cmex10 scaled \magstep1 \font\twelveit=cmti12
\font\tenss=cmss10
 
 \font\ninebf=cmbx9
\font\ninerm=cmr9 \font\ninei=cmmi9
\font\ninesy=cmsy9 
\font\eightrm=cmr8
\catcode`@=11
\newskip\ttglue
\newfam\ssfam

\def\twelvepoint{\def\rm{\fam0\twelverm}
\textfont0=\twelverm \scriptfont0=\ninerm \scriptscriptfont0=\sevenrm
\textfont1=\twelvei \scriptfont1=\ninei \scriptscriptfont1=\seveni
\textfont2=\twelvesy \scriptfont2=\ninesy \scriptscriptfont2=\sevensy
\textfont3=\twelveex \scriptfont3=\twelveex \scriptscriptfont3=\twelveex
\def\it{\fam\itfam\twelveit} \textfont\itfam=\twelveit
\def\sl{\fam\slfam\twelvesl} \textfont\slfam=\twelvesl
\def\bf{\fam\bffam\twelvebf} \textfont\bffam=\twelvebf
\scriptfont\bffam=\ninebf \scriptscriptfont\bffam=\sevenbf
\def\tt{\fam\ttfam\twelvett} \textfont\ttfam=\twelvett
\def\ss{\fam\ssfam\twelvess} \textfont\ssfam=\twelvess
\tt \ttglue=.5em plus .25em minus .15em
\normalbaselineskip=14pt
\abovedisplayskip=14pt plus 3pt minus 10pt
\belowdisplayskip=14pt plus 3pt minus 10pt
\abovedisplayshortskip=0pt plus 3pt
\belowdisplayshortskip=8pt plus 3pt minus 5pt
\parskip=3pt plus 1.5pt
\setbox\strutbox=\hbox{\vrule height10pt depth4pt width0pt}
\let\sc=\ninerm
\let\big=\twelvebig \normalbaselines\rm}
\def\twelvebig#1{{\hbox{$\left#1\vbox to10pt{}\right.\n@space$}}}

\def\tenpoint{\def\rm{\fam0\tenrm}
\textfont0=\tenrm \scriptfont0=\sevenrm \scriptscriptfont0=\fiverm
\textfont1=\teni \scriptfont1=\seveni \scriptscriptfont1=\fivei
\textfont2=\tensy \scriptfont2=\sevensy \scriptscriptfont2=\fivesy
\textfont3=\tenex \scriptfont3=\tenex \scriptscriptfont3=\tenex
\def\it{\fam\itfam\tenit} \textfont\itfam=\tenit
\def\sl{\fam\slfam\tensl} \textfont\slfam=\tensl
\def\bf{\fam\bffam\tenbf} \textfont\bffam=\tenbf
\scriptfont\bffam=\sevenbf \scriptscriptfont\bffam=\fivebf
\def\tt{\fam\ttfam\tentt} \textfont\ttfam=\tentt
\def\ss{\fam\ssfam\tenss} \textfont\ssfam=\tenss
\tt \ttglue=.5em plus .25em minus .15em
\normalbaselineskip=12pt
\abovedisplayskip=12pt plus 3pt minus 9pt
\belowdisplayskip=12pt plus 3pt minus 9pt
\abovedisplayshortskip=0pt plus 3pt
\belowdisplayshortskip=7pt plus 3pt minus 4pt
\parskip=0.0pt plus 1.0pt
\setbox\strutbox=\hbox{\vrule height8.5pt depth3.5pt width0pt}
\let\sc=\eightrm
\let\big=\tenbig \normalbaselines\rm}
\def\tenbig#1{{\hbox{$\left#1\vbox to8.5pt{}\right.\n@space$}}}
\let\rawfootnote=\footnote \def\footnote#1#2{{\rm\parskip=0pt\rawfootnote{#1}
{#2\hfill\vrule height 0pt depth 6pt width 0pt}}}
\def\tenfoot{\tenpoint\hskip-\parindent\hskip-.1cm}
\def\ft#1#2{{\textstyle{{#1}\over{#2}}}}
\twelvepoint
\def\fish{\kern-.25em{\phantom{abcde}\over \phantom{abcde}}\kern-.25em}

\def\lagr{{\cal L}}

\def\noverm#1#2{{\textstyle{#1\over #2}}}
\def\half{\noverm{1}{2}}


\def\ft#1#2{{\textstyle{{#1}\over{#2}}}}
\def\frac#1#2{{{#1}\over{#2}}}
\def\1#1{\frac1{#1}} \def\2#1{\frac2{#1}} \def\3#1{\frac3{#1}}

\def\sb#1{\lower.4ex\hbox{${}_{#1}$}}

\def\Vt{{\widetilde V}}

\def\.{\,\,,\,\,}
\def\FF#1#2#3#4#5{\,\sb{#1}F\sb{\!#2}\!\left[\,{{#3}\atop{#4}}\,;{#5}\,\right]}

\def\cramp{\medmuskip = 2mu plus 1mu minus 2mu}

\cramp


\def\a{\alpha}

\def\t{\theta}

\def\f{\phi}

\def\lagr{{\cal L}}

\def\noverm#1#2{{\textstyle{#1\over #2}}}
\def\half{\noverm{1}{2}}


\oneandahalfspace
\vskip 2truecm
\rightline{IC/91/206}
\rightline{CTP TAMU-9/91}
\vskip 1.5cm
\centerline{\bf ASPECTS of  $W_{\infty}$ SYMMETRY\footnote{$^*$}{To appear
in the proceedings of the 4th Regional Conference on Mathematical Physics,
Tehran, Iran, 1990}}
\vskip 1.5truecm \centerline{Ergin Sezgin}
\vskip 1truecm
\centerline{International Center for Theoretical Physics, 34100 Trieste,
Italy\footnote{$^\dagger$}{Permanent Address: Center for Theoretical Physics,
Texas A\& M University, College Station, TX 77843, USA.}}
\vskip 2truecm

\centerline{ABSTRACT}
\bigskip
{ We review the structure of $W_\infty$ algebras, their super and
topological extensions, and their contractions down to (super)
$w_\infty$. Emphasis is put on the field theoretic realisations of these
algebras. We also review the structure of $w_\infty$ and $W_\infty$ gravities
and comment on various applications of $W_\infty$ symmetry.}

\vfil\eject

\bigskip \centerline{\bf 1.Introduction}
\bigskip

$W_{\infty}$ algebra is a particular generalization of the Virasoro algebra
which contains fields of conformal spins $2,3,...,\infty$ [1]. Virasoro algebra
is a special truncation of $W_\infty$ algebra containing only the spin-2 field:
the 2D energy momentum tensor. The importance of Virasoro algebra in string
theory is well known. Virasoro algebra also arises as a
symmetry group of the KdV (Korteveg-de Vries) hierarchy which, in turn, plays
an
important role in the study of non-perturbative 2D gravity coupled to minimal
conformal matter [2].

        It is natural to search for a higher
spin generalization of the Virasoro algebra which may lead to a  $W$-string
theory containing {\it massless} higher spin fields in {\it target spacetime}
[3]. This would mean a dramatic enlargement of the usual Yang-Mills  and
diffeomorphism symmetries in spacetime. A more immediate
application of $W_\infty$ symmetry is based on the fact that it arises as a
symmetry group of the KP (Kadomtsev-Petviashvili) hierarchy [4,5], which is
expected to be closely associated with 2D gravity coupled to a $c=1$
matter, i.e. string in one dimension. Furthermore, a certain contraction of
$W_\infty$, known as $w_\infty$ [6,7] emerges as a symmetry group in the study
of self-dual gravity, which can be formulated as the  $sl(\infty)$ Toda theory
[8], as well as the natural background for the critical $N=2$ superstring [9].

          In this review, we begin by recalling the field theoretic realisation
of
the Virasoro algebra. We then outline the generalisation of the construction
for $W_\infty$ and $W_{1+\infty}$ algebras [10]. For completeness we also
describe the super $W_\infty$ [11] as well as the topological $W_\infty$
algebras [12]. Contraction of super $W_\infty$ down to super $w_\infty$ algebra
is discussed next. We then decsribe the analog of
the 2D Lagrangian ${\cal L}= {1\over 2}{\sqrt -h}h^{ij}\partial_i\phi
\partial_j\phi$ (in a gauge in which the Weyl symmetry is fixed) which has
local $w_\infty$ symmetry, and is referred to as $w_\infty$ gravity. A similar
construction of $W_\infty$ gravity [14] is also reviewed. We conclude by
summarising some known facts about various applications of $W_\infty$
symmetry.

\bigskip
\centerline{\bf 2. Virasoro\ Algebra}
\bigskip
Let us recall how the Virasoro algebra arises in the context of a 2D
conformal  field
theory of a free complex scalar $\varphi$. The normal ordered energy momentum
tensor is given by
$$
   \eqalign{
                T(z)&=-:\partial\varphi(z)\partial\varphi^*(z):  \cr
                                                      &\equiv
-\lim_{z\rightarrow w}
\bigg[\partial\varphi(z)\partial\varphi^*(\omega) +{1\over (z-w)^2}\bigg].\cr}
\eqno(1)
$$
The conformal spin $j$ of a (chiral) field $\Phi(z)$ is defined by
$$
T(z) \Phi(w) \sim {j\Phi\over(z-w)^2}+{\partial \Phi\over z-w}. \eqno(2)
$$
Using the two-point function
$$
         \varphi(z) \varphi^*(w)\sim -log(z-w),  \eqno(3)
$$
a simple computation involving the use of Wick rules and Taylor expansion
yield
   s the following
well known operator product expansion (OPE):
$$
 T(z)T(w)\sim {\partial T\over z-w}+{2T\over (z-w)^2}+{1\over(z-w)^4}.\eqno(4)
$$
Comparing with (2), we see that $T(z)$ indeed has conformal spin 2. In terms of
the Fourier modes defined by
$$
 L_n=\oint_C {dz\over 2\pi i}z^{n+1} T(z), \eqno(5)
$$
one deduces the Virasoro algebra with cental extension $c=2$:
$$
    [L_m,L_n]= (m-n)L_{m+n}+{1\over 12}(m^3-m)\delta_{m+n,0}. \eqno(6)
$$
(For a detailed description of this calculation, and of conformal field
theories
    in general,
see the excellent review by Ginsparg [15]). The associativity of OPE garanties
that the Jacobi identities are satisfied by this algebra. Replacing the last
coefficient ${1\over 12}$ by ${c\over 24}$ one obtains the Virasoro algebra
with central extension $c$. The Jacobi identities are no longer garantied, but
one checks that they are indeed still satisfied. At this stage, we may treat
the  Virasoro algebra in its own right, forgeting about the specific
realisation which lead to it. Of course, there are {\it many} ways of arriving
at the Virasoro algebra. We have chosen to describe it through a field
theoretic realisation, because this furnishes an intuitive way of understanding
higher spin generalizations, as we shall see below.
\bigskip
\centerline {\bf 3. $W_\infty$ \ and\ $W_{1+\infty}$\ Algebras}
\bigskip
        A generalization of the Virasoro algebra containing a spin-3 generator
w
as
discovered in [16]. It
is a nonlinear algebra since the commutator of two spin-3 generators yields,
among other
terms, a term which is quadratic in spin-2 generator. This is known as the
$W_3$ algebra. A further
generalization to $W_N$ algebra where generators with integer spins
$2,3,...,N$ occur was found in
ref.~[17]. There are various inequivalent ways of taking an $N\rightarrow
\infty$ limit of $W_N$. A very simple one leading to a linear algebra, called
$w_\infty$ algebra was found in ref.~[6]. All the nonlinearities of the $W_N$
are washed away in this algebra, and it is indeed an infinite dimensional Lie
algebra. Geometrical interpretation of $w_\infty$ is that it is the algebra of
symplectic-diffeomorphisms of a cylinder [1], as we shall see later.

A characteristic feature of $w_\infty$ is that the commutator of a spin-i
genera
   tor with a spin-j
generator yields a spin $(i+j-2)$ generator alone, and the only possible
central
    extension is in the
Virasoro sector. A nontrivial generalization of it, called the $W_\infty$
algebra, in which central terms arise in all spin sectors, and in which the
commutator of a  spin-i generator with a spin-j generator yields generators
with spins $(i+j-2)$, $(i+j-4)$,..., down to spin-2, was discovered in
ref.~[1].
An important feature of this algebra, besides containing the central charges in
all spin sectors, is that it is indeed an infinite dimensional Lie algebra,
since all the non-linearities of the $W_N$ algebra are absent, just as in the
case of $w_\infty$. The construction of this algebra has been
discussed in great detail in ref.~[1].  Below, we  shall motivate it by
considering its field theoretical realisation, as we did for the Virasoro
algebra in the previous section.

A natural way to generalize the construction of the previous section is to
intro
   duce higher derivative
analogs of (1) as follows [18]:
$$
V^i(z)=\sum_{k=0}^{i}
\alpha_k^i:\partial^{k+1}\varphi(z)\partial^{i+1-k}\varphi^*(z):, \eqno(7)
$$
where
$\alpha_k^i$ are for the moment arbitrary constant coefficients.  Choosing
$\alpha^0_0=-1$, yields an expression for $V^0(z)$ which coincides with the
energy momentum tensor $T(z)$. The next bilinear current contains terms of the
form
$$
   V^1:\quad\quad\quad :\partial \varphi \partial^2 \varphi^*: ,\qquad
:\partial
   ^2 \varphi \partial
\varphi^*:
$$
The sum of the two terms can be written as the derivative of the lower spin
curr
   ent $V^0$, but the
difference is an independent combination. Similarly, at the next level we have
$$
V^2: \quad\quad\quad :\partial\varphi\partial^3\varphi^*:,\qquad
:\partial^2\varphi\partial^2\varphi^*:, \qquad
:\partial^3\varphi\partial\varphi
   ^*:.
$$
With a little effort one can show that one linear combination can be written as
   the derivative
the independent $V^1$ and another linear combination can  be written as the
doub
   le derivative of
$V^0$. Thus, one is left with a single independent linear combination at this
le
   vel. This
pattern continues at higher levels: At level $i$ there are $(i+1)$ possible
bili
   near terms
and $i$ combination of them can be written as derivatives of lower level
terms, while only one combination is independent. This means that, with any
nondegenerate choice of the coeffecients $\alpha^i_k$ the currents $V^i$ are
garantied to give closed OPE algebra.   By inspection we can see that the OPE
product of $V^i$ with $V^j$ , will produce a sum of $V^{i+j-k}$, for
$k=2,4,...,i+j$ the last term corresponding to the central extension. Defining
the Fourier modes $$
      V^{\ell}_m=\oint_C {1\over 2\pi i} V^{\ell}(z) z^{m+\ell +1}, \eqno(8)
$$
one finds an algebra  of the form $[V^i_m, V^j_n]\sim
V^{i+j}_{m+n}+V^{i+j-2}_{m
   +n}+...$, with
appropriate coefficients, with``spin'' decreasing all the way down to zero. (We
shall be more precise about the conformal spin of the fields $V^i(z)$ below).
The last term, which is the central extension, will turn out to have a specific
value due to the specific choice of realisation involving a single complex
scalar. As before, assigning an arbitrary value to the central extension, one
finds that the Jacobi identities are still satisfied by the algebra. Thus, as
in the case of Virasoro algebra, we have an infinite dimensional algebra in its
own right, regardless of the particular realisation that lead to it. The
algebra obtained in this manner is the $W_\infty$ algebra [1].

Determining the structure constants is by no means an easy task. One guiding
consideration is to work in a basis in which the central term is diagonal in
spi
   n, i.e.
nonvanishing only between any pair of equal spin generators. It turns out that,
   this
requirement fixes the coeffients $\alpha^i_m$ upto an overall i-dependent
scalin
   gs.  With a particular
choice of such an overall scaling these coefficients are given by [18]
$$
\a^i_k= \frac{(-1)^{k+1}
2^{-i-1}(i+2)!}{(2i+1)!!(i+1)}{i+1\choose k}{i+1\choose k+1}, \eqno(9)
$$ and the $W_\infty$ algebra takes the form [1]
$$
 [V_m^i,V_n^j] =\sum_{\ell\ge 0}\, g_{2\ell}^{ij}(m,n)\,
V_{m+n}^{i+j-2\ell}+c_i(m)\delta^{ij}\delta_{m+n,0}, \eqno(10)
$$
where
$$
c_i(m)= c\,{2^{2i-3}i!(i+2)!\over (2i+1)!!(2i+3)!!}\prod_{k=-i-1}^{i+1} (m+k)
\eqno(11)
$$

Although it is helpful to examine the OPE products of the currents $V^i(z)$ to
d
   etermine the structure
constants $g^{ij}_{2\ell}$, it is more advantageous to directly impose the
Jacob
   i identities on the
algebra (10). In fact, that is what was originally done in ref.~[1], where it
was found that
$$
g_{\ell}^{ij}(m,n) ={1\over
2(\ell+1)!}\,\phi_{\ell}^{ij}\, N_{\ell}^{ij}(m,n), \eqno(12)
$$
where
$$
  \eqalign{
 \phi^{ij}_{\ell} &=
\FF43{-\ft12\.\ft32\.-\ell-\ft12\.-\ell}{-i-\ft12\.-j-\ft12\.i+j-2\ell+\ft52}1,
\cr
N^{ij}_{\ell}(m,n) &=
\sum_{k=0}^{\ell+1}(-1)^k{\ell+1\choose k}
[i+1+m]_{\ell+1-k}[i+1-m]_k[j+1+n]_k[j+1-n]_{\ell+1-k}. \cr}\eqno(13)
$$
and $[a]_n\equiv a!/(a-n)!$, while the hypergeometric function $_4F_3 (z)$
is defined by
$$
\FF43{a_1\.a_2\.a_3\.a_4}{b_1\.b_2\.b_3}z
=\sum_{n=0}^{\infty}{(a_1)_n(a_2)_n(a_
   3)_n(a_4)_n\over
 (b_1)_n(b_2)_n(b_3)_n} {z^n\over n!}, \eqno(14)
$$
where $(a)_n\equiv (a+n-1)!/(a-1)!$. The sum terminates if any one of the
parameters $a_1,...,a_4$ is zero or a negative integer. The function
$N_{\ell}^{ij}(m,n)$ are related to the SL(2,R) Clebsch-Gordan coefficients,
while $\phi^{ij}_{\ell}$ are formally related to Wigner 6-j symbols [1].

It is worth mentioning that while $V^0$ and $V^1$ correspond to true primary
fields of conformal spin 2 and 3, respectively,  $V^i$ for $i\ge 2$ correspond
to {\it quasi-primary} fields of spin $i+2$ since lower spin terms occur in the
   commutator $[V^0_m,
V^i_n]$ in that case.The choice of basis is relevant to this fact. In another
basis, $V^i(z)$ may not be even quasi-primary. Also, one can define primary
fields for all $i\ge 2$ as well, but then the algebra becomes nonlinear [19].

  The $W_{\infty}$ algebra can be enlarged to contain a
spin-1 generator. The resulting algebra is called $W_{1+\infty}$ algebra,
and has the following form [10]
$$
[\Vt_m^i,\Vt_n^j] =\sum_{\ell=0}^{\infty}\,{\tilde g}_{2\ell}^{ij}(m,n)\,
\Vt_{m+n}^{i+j-2\ell}\, +\, {\tilde c}_i(m)\delta^{ij}\delta_{m+n,0}, \eqno(15)
$$
where
$$
 {\tilde c}_i(m) =
c\,{2^{2i-3}((i+1)!)^2\over(2i+1)!!(2i+3)!!}\prod_{k=-i-1}^{i+1} (m+k)
\eqno(16)
$$
and the structure constants are
$$
{\tilde g}_{\ell}^{ij}(m,n) ={1\over
2(\ell+1)!}\,{\tilde \phi}_{\ell}^{ij}\, N_{\ell}^{ij}(m,n), \eqno(17)
$$
with $N_{\ell}^{ij}(m,n)$ given as before, and
$$
     {\tilde \phi}^{ij}_{\ell} =
\FF43{-\ft12\.\ft12\.-\ell-\ft12\.-\ell}
          {-i-\ft12\.-j-\ft12\.i+j-2\ell+\ft52}1.  \eqno(18)
$$
There exists a redefinition of the generators, involving finite number of
terms at each spin level, which enables one to truncate the $W_{1+\infty}$
algebra down to the $W_{\infty}$ subalgebra [20]. Another
interesting subalgebra of the $W_{1+\infty}$ algebra, denoted by
$W^+_{1+\infty}$, is obtained by restricting the Fourier modes of the
generators $V^\ell_m$ to lie in the range $m\ge -\ell-1$.

Currents obeying the $W_{1+\infty}$ algebra can be
constructed in terms of a complex fermion $\psi(z)$ with the propagator
$$
\psi^*(z)\psi(w)\sim{1\over{z-w}}, \eqno(19)
$$
as follows [21,22,11]
$$
\Vt^i(z) =\sum_{k=0}^{i+1} \beta^i_k :\partial^k\psi^*\partial^{i+1-k}\psi:,
\eqno(20)
$$
where the coefficients $\beta^i_k$ are given by
$$
\beta_k^i = (-1)^k\frac{(i+1)(k+1)}{(i+2)(i-k+1)}\alpha^i_k, \eqno(21)
$$
with $\alpha^i_k$ as defined in (9). The currents $\Vt^i(z)$ have (quasi-)
conformal spins $s=i+2$.

{}From the above realization of $W_{1+\infty}$ in terms
of fermionic field $\psi(z)$, one can obtain a bosonic realisation of it in
terms of a single scalar field $\phi(z)$, through the bosonization formulae
$$
    \eqalign{
         \psi(z) &= : e^{\phi(z)} : \ \ \cr
          \psi^*(z) &= : e^{-\phi(z)} :  \cr}\eqno(22)
$$
The result is [23]
$$
      {\tilde V}^i (z)=\sum_{\ell=0}^{i+1}(-1)^\ell
{([i+1]_\ell)^2\over (i-\ell+1)\ell![2i]_\ell}\partial^\ell
P^{(i-\ell+1)}(z)  \eqno(23)
$$
where
$$
      P^{(k)}(z) \equiv : e^{-\phi(z)}\partial^k e^{\phi(z)} : \eqno(24)
$$
First few examples are
$$
\eqalign{
       {\tilde V}^{-1} &= \partial \phi  \cr
    {\tilde V}^0 &= {1\over 2} : (\partial \phi)^2 :  \cr
  {\tilde V}^1 &= {1\over 3} : (\partial \phi)^3 :   \cr
     {\tilde V}^2 &=  {1\over 4} : (\partial \phi)^4 : -{3\over 20}:
(\partial^2
\phi)^2 :  +{1\over 10} : \partial \phi \partial^3 \phi :  \cr}\eqno(25)
$$
\bigskip
\centerline{\bf 4. Super-$W_\infty$ \ Algebra}
\bigskip
There exists  an $N=2$ supersymmetric extension of $W_\infty$
algebra [11]. More precisely, it is the superextension of the bosonic algebra
$W_\infty \oplus W_{1+\infty}$, in which the following supercurrents are
introduced [11]
$$
G^{\alpha} =
\sum_{k=0}^\a \gamma^\a_k :\partial^{\a +1-k}\f^*\partial^k\psi:,\eqno(26)
$$
where
$$
     \gamma_k^i = (-1)^k \frac{2(k+1)}{(i+2)}\alpha^i_k, \eqno(27)
$$
with $\alpha^i_k$ as defined in (9). The currents $G^\alpha$ have (quasi-)
conformal spins $s=\alpha+\ft32$.

One finds that the currents (7), (20) and (26)
obey the super-$W_\infty$ algebra, which, in terms of the Fourier modes
of the generators, is given by (10), (15) and the following relations [11]
$$
\eqalign{
    \{ {\bar G}^\alpha_r, G^\beta_s \} &=\sum_{\ell =0}^{\infty}\,
\Big(b_\ell^{\alpha \beta} (r,s)\, V_{r+s}^{\alpha +\beta-\ell} \,+\,{\tilde
b}_\ell^{\alpha\beta}(r,s)\, \Vt_{r+s}^{\alpha+\beta-\ell} \Big)\,+\,
{\hat c}_\alpha(r)\delta^{\alpha\beta}\delta_{r+s,0}, \cr
[V_m^i, G_r^\alpha ]&= \sum_{\ell=0}^\infty\,
a_\ell^{i\alpha}(m,r)\,G_{m+r}^{\alpha+i-\ell +1},  \cr
[\Vt_m^i, G_r^\alpha ]&= \sum_{\ell=0}^\infty\,
{\tilde a}_\ell^{i\alpha}(m,r)\,G_{m+r}^{\alpha+i-\ell +1},  \cr
[V_m^i, {\bar G}_r^\alpha ]&= \sum_{\ell=0}^\infty \,
(-1)^{\ell +1} a_\ell^{i\alpha}(m,r)\,{\bar G}_{m+r}^{\alpha+i-\ell +1}, \cr
[\Vt_m^i, {\bar G}_r^\alpha ]&= \sum_{\ell=0}^\infty\,
(-1)^{\ell +1} {\tilde a}_\ell^{i\alpha}(m,r)\, {\bar G}_{m+r}^{\alpha +i-\ell
+1}.  \cr} \eqno(28)
$$
Explicit expressions for ${\hat c}_\alpha(r)$ and the structure constants
${\tilde g}, \ldots ,{\tilde a}$ can be found in ref.~[11]. The spectral flow
in
the above algebra, as well as its various truncations including the  $N=1$
supersymmetric truncation is also discussed in ref. ~[11].

One can choose other types of higher derivative bilinear currents  giving  rise
   to a
super-$W_{\infty}$ algebra which isomorphic to the above one. For example, one
c
   an choose
$\Vt^i(z)\sim \psi^* \partial^{i+1}\psi$. Although this gives rise to central
te
   rms between
generators of different spin, this realisation may nontheless have  some
advanta
   ges in dealing with
certain  problems [5,24].

The redefinition of the basis elements in $W_{1+\infty}$ algebra has been
discussed in ref.~[20],
where it was shown that a one paramater family of isomorphic algebras can be
obt
   ained in this way. One
application of basis redefinition is to show that the spin-1 generator can be
co
   nsistently truncated,
as a consequence of which one is left with $W_\infty$ algebra. Recently, these
i
   ssues have been
treated in great detail in ref.~[25], where a one parameter family of
isomorphic
super $W_\infty$ algebras, denoted by super $W_{\infty}(\lambda)$, were also
obtained. The construction of ref.~[25] uses an infinite set of differential
operators, and enables them to compute explicitly the structure constants of
the algebra for any value of the parameter $\lambda$. These generators are of
the form [25]
$$
V^i_\lambda(\Omega^{(i)})=\sum_{k=0}^{2i+2}A^k(i,\lambda)
\big(D^{2i+2-k}\Omega^{(i)}\big) D^k,   \eqno(29)
$$
where $i=0,\half,1,...$, $\Omega^{(i)}(z,\theta)$ are the generating
superfunctions, $D={\partial\over \partial\theta}-\theta\partial$ and the
coefficients $A^k(i,\lambda)$ can be found in ref.~[25].

A geometric interpretation of  $W_{1+\infty}$ is that it is the algebra of
smooth differential operators on a circle [20]. It can also be viewed as the
universal enveloping algebra of the $U(1)$ Kac-Moody algebra generated by
$\Vt^{-1}_m\equiv j_m$ modulo the ideal $j_mj_n-j_{m+n}=0$ [20]. Similarly,
$W_{\infty}$ can be viewed as the  universal enveloping algebra of the Virasoro
algebra modulo the ideal generated by $L_m L_n-(L_0 +m) L_{m+n}=0$ [20]. There
is also an underlying $SL(2,R)$ structure. One can first built the universal
enveloping algebra of $SL(2,R)$, denoted by $U_L(SL(2,R))$, in which the
generators are labeled by $SL(2,R)$ spin j and ``magnetic quantum number''
$m\le |j|$. One can then extend the range of $m$ to $-\infty <m<\infty$ in the
structure constants [1]. This procedure typically gives an algebra with all
conformal spins $-\infty <s<\infty$. Remarkably, only in the case when one
modes out the $U_L(SL(2,R))$ by the ideal  $C_2=0$[1] or $C_2+{1\over 4}=0$
[10], where $C_2$ is the second Casimir operator of $SL(2,R)$, that one obtains
an infinite dimensional algebra in which only positive spin generators occur.
In
fact, those are precisely the $W_\infty$ and $W_{1+\infty}$ algebras,
respectively. Dramatic simplifications occur in the structure constants of
$U_L(SL(2,R))$ algebra moded by the ideal $C_2+{3\over 16}=0$. However, this
algebra, associated with the so called symplecton, singleton or metaleptic
representation of $SL(2,R)$, does not admit extension to a $W$-like algebra  in
which generators of a given spin would correspond to a (quasi) primary
conformal field [1].

It would be interesting to apply the above considerations to the case of
univers
   al enveloping
algebras of any Lie (super) algebra. In this context, see the interesting work
of Fradkin and Linetsky [26]. See also ref.~[27], where an SU(N) structure is
introduced by adding new generators $V^{i,a}_n$ where the index
$a=1,...,p^2-1$ labels the adjoint representation of $SU(p)$. The generator
$V^{-1,a}_n$ obeys the affine ${\hat {su}}(p)$ algebra with level $k$. The
$N=2$ supersymmetric  version of this algebra, contains the generators $V^i_n$,
$\Vt^i_n$, $V^{i,a}_n$, $G^{\alpha,A}_r$ and ${\bar G}^{\alpha,A}_r$, where the
index $A=1,...,p$ labels the fundamental representation of $SU(p)$. In the
bosonic case, $V^i$ and $V^{i,a}$ can be combined to represent the $p^2$
dimensional representation of $U(p)$, denoted by $V^{i,r},\ (r=0,1,...,p^2-1)$,
and the currents (7) are now replaced by
$$
V^{i,r}(z)=\sum_{A,B=1}^{p}\sum_{k=0}^{i} \alpha_k^i T^r_{AB}
:\partial^{k+1}\varphi^A(z)\partial^{i+1-k}\varphi^{*B}(z):,  \eqno(30)
$$
where $T^r_{AB}$ are the generators of $U(p)$ in the fundamental
representation. The algebra generated by these currents is called
$W_\infty^p$, and it can be viewed as the large N limit of the Grassmannian
coset model $G_N(p)=SU(p+1)_n/SU(p)_N\otimes U(1)$ [28]. The large p limit of
$W_\infty^p$ takes the form [28]
$$
[V_m^{i,\vec k},
V_n^{j,\vec \ell}]=[(j+1)m-(i+1)n]
V^{i+j,{\vec k+\vec \ell}}_{m+n}+
{\vec k}\times {\vec \ell}\ V^{i+j+1,{\vec k+\vec \ell}}_{m+n}, \eqno(31)
$$
where ${\vec k}=(k_1,k_2)$ and ${\vec \ell}=(\ell_1,\ell_2)$. One can show that
this algebra describes symplectic diffeomorphisms in four dimensions [28].

\bigskip
\centerline{\bf 5. Topological\ $W_\infty$ \ Algebra}
\bigskip
The $N=2$ super-$W_{\infty}$ algebra described above
can be twisted to give a topological algebra which we call $W_{\infty}^{\rm
top}$ [12]. The idea is to identify one of the fermionic generators as the
nilpotent BRST charge Q, and to define  bosonic generators which can be written
in the form ${\hat V}^i_m =\{ Q,{\rm something} \}$. This is the higher-spin
generalization of the property that holds for the energy-momentum tensor of a
topological field theory. A suitable candidate for the BRST charge is
$$
                Q=-{\bar G}^0_{-{1\over 2}}.\eqno(32)
$$
Inspired by the twisting of the $N=2$ super-Virasoro algebra, we then define
the
    generators of
$W_{\infty}^{\rm top}$ to be $G^i_{m+{1\over 2}}$ and define ${\hat V}^i_m$ as
follows
$$
{\hat V}^i_m=-\{Q,G^i_{m+{1\over2}}\}.
 \eqno(33)
$$
It can be easily shown that these generators form a closed algebra. Finding the
   structure constants
requires more work. Towards that end, from (33) we first find
$$
{\hat V}^i_m=V^i_m+{\tilde V}^i_m -{2i(m+i+1)\over 2i+1} V^{i-1}_m +{(2i+2)
(m+i+1)\over 2i+1}{\tilde V}^{i-1}_m. \eqno(34)
$$
With respect to the new energy-momentum tensor ${\hat V}^0(z)$ the
(quasi)confor
   mal spin of the
fields ${\hat V}^i(z)$ is shifted up by a half.  After some algebra, we then
fin
   d that the $W^{\rm
top}_\infty$ algebra takes the form [12]
$$
\eqalign{
[{\hat V}^i_m,{\hat V}^j_n]&=\sum_{\ell\ge0}{\hat g}^{ij}_\ell(m,n) {\hat
V}^{i+j-\ell}_{m+n},\cr
[{\hat V}^i_m,G^j_{n+\ft12}]&=\sum_{\ell\ge0} {\hat g}^{ij}_\ell(m,n)
G^{i+j-\ell}_{m+n+\ft12},\cr
\{G^i_{m+\ft12},G^j_{n+\ft12}\}&=0,\cr}\eqno(35)
$$
where
$$
\eqalign{
{\hat g}^{ij}_\ell(m,n)&=a^{ij}_{\ell+1}(m,n+\ft12) +{\tilde a}^{ij}_{\ell
+1}(m,n+\ft12)\cr
& -{2i(m+i+1)\over 2i+1} a^{i-1,j}_\ell(m,n+\ft12) +{(2i+2)
(m+i+1)\over 2i+1} {\tilde a}^{i-1,j}_\ell(m,n+\ft12).\cr}\eqno(36)
$$

Note that the structure constants for $[{\hat V}^i_m,G^j_{n+\ft12}]$ are the
same as those for $[{\hat V}^i_m,{\hat V}^j_n]$, and that the algebra is
centerl
   ess. Furthermore, one
can show that ${\hat V}^i_m$'s generate the diagonal subalgebra in
$W_{\infty}\t
   imes W_{1+\infty}$
with vanishing central charge. A field theoretic realisation of
$W^{\rm top}_{\infty}$ is given in ref.~[12].
\bigskip
\centerline{\bf 6. Contraction\ of\ $W_\infty$\ Down\ to\ $w_\infty$}
\bigskip
There exists an interesting contraction of all the $W$ algebras discussed
above.
    Consider for
simplicity $W_{1+\infty}$ algebra. A rescaling  of the form $\Vt^i_m\rightarrow
   q^{-i}v^i_m$ followed
by the limit $q\rightarrow 0$ yields the following remarkably simple result:
$$
[v^i_m, v^j_n] = [(j+1)m-(i+1)n] v^{i+j}_{m+n}+{c\over
12}(m^3-m)\delta^{i,0}\delta^{j,0}\delta_{m+n,0}, \eqno(37)
$$
where $v^i_m$ now represent the Fourier modes of a true primary field of
conform
   al spin $(i+2)$,
and $i\ge -1$ and $-\infty <m<\infty$. This is known as the $w_\infty$ algebra.
Note that all the lower spin terms on the right hand side are now absent, and
that all the central charges have also disappeared except in the Virasoro
sector. The above algebra has the nice geometric interpretation [6] as the
symplectic diffeomorphisms of a cylinder [1]. (In two dimensions symplectic
diffeomorphisms coincide with are area-preserving diffeomorphisms). In general,
the symplectic diffeomorphisms of a $2n$ dimensional symplectic manifold are
those which leave the symplectic structure (a non-degenerate 2-form) invariant,
and they are easily shown to be generated by vector fields of the form
$\xi^a=\Omega^{ab}\partial_b\Lambda$, where $\Omega^{ab}$ are the components of
the inverse of the symplectic 2-form, and $\Lambda$ is an arbitrary function.
The generator of the symplectic-diffeomorphisms can be written as
$$
        \eqalign{
              v_\Lambda &=\xi^a(\Lambda) \partial_a  \cr
                        &=\Omega^{ab}\partial_b\Lambda\partial_a \cr}\eqno(38)
$$
These generators obey the algebra $[v_{\Lambda_1},
v_{\Lambda_2}]=v_{\Lambda_3}$
    where the composition
parameter $\Lambda_3$ is given by the Poisson bracket $\Lambda_3
=\Omega^{ab}\partial_b\Lambda_1\partial_a\Lambda_2$. Let us now consider the
cas
   e of a
cylinder and define a complete set of functions on the cylinder
$S^1\times R$ with coordinates $0\le x\le 2\pi$ and $-\infty\le y\le \infty$
as follows
$$
                     u^\ell_m= -ie^{imx}y^{\ell+1} \eqno(39)
$$
Expanding $\Lambda(x,y)= \sum_{\ell,m}^{} \Lambda^\ell_m u^\ell_m$, and
defining the Fourier modes of the generators as
$v_\Lambda=\sum_{\ell,m}\Lambda^\ell_m v^\ell_m$, one finds that
$$
v^\ell_m= \Omega^{ab}\partial_b u^\ell_m \partial_a \eqno(40)
$$
It can be easily shown that these generators obey the centerless part of the
$w_\infty$ algebra (37). In ref.~[29], on quite general grounds, it was found
that the symplectic-algebra of a Riemann surface of genus $g$ admitted a $2g$
parameter central extension. Since the cylinder has genus one, we expect that
the algebra admits a one parameter central extension. From the formula in
ref.~[29], one can check that it indeed has the form given in (37).
                                                We next consider the
contraction
 of the $N=2$ super $W_\infty$ algebra
down to the corresponding super $w_\infty$ algebra. To this end we perform the
rescalings
$$
   \eqalign{
               V^\ell_m &\rightarrow q^{-\ell}(v^\ell_m -
            {1\over 2}q^{-1}J^\ell_m)/2,  \cr
               {\tilde V}^\ell_m &\rightarrow q^{-\ell}(v^\ell_m +
{1\over 2}q^{-1}J^\ell_m)/2,  \cr
          G^\alpha &\rightarrow q^{-\ell} G^\alpha{\sqrt 2}, \cr
         {\bar G}^\alpha &\rightarrow q^{-\ell} {\bar G}^\alpha{\sqrt 2}.\cr}
          \eqno(41)
$$
Taking the limit $q\rightarrow 0$ in the $N=2$ super $W_\infty$
algebra  given in (10), (15) and (28) yields the result
[30]\footnote{$^\dagger$}{\tenfoot To obtain this result, it is sufficient to
know that $g^{ij}_0={\tilde g}^{ij}_0= (j+1)m-(i+1)n,\
b^{\alpha\beta}_0={\tilde b}^{\alpha\beta}_0=1,\
b_1^{\alpha\beta}(r,s)=-{\tilde
b}^{\alpha\beta}_1(r,s)=[(2\beta+1)r-(2\alpha+1)s], \ a^{i\alpha}_0=-{\tilde
a}^{i\alpha}_0={-1\over 4},\ a_1^{i\alpha}(m,r)={\tilde a}_1^{i\alpha}(m,r)=
{1\over 4}[(2\alpha+1)m -(2i+2)r].$}
$$
\eqalign{
   [v^i_m, v^j_n] &=\big[ (j+1)m-(i+1)n\big]
v^{i+j}+{c\over 8}(m^3-m) \delta^{i,0}\delta^{j,0}\delta_{m+n,0}, \cr
   [v^i_m, J^{j-1}_n] &= \big[jm-(i+1)n\big] J^{i+j-1}_{m+n}, \cr
\{{\bar G}^\alpha_r, G^\beta_s \} &=2 v^{\alpha +\beta}_{r+s}
-2\big[(\beta+\half )r-(\alpha+\half)s\big]J^{\alpha+\beta-1}_{r+s}+{c\over
2}(r^2-{1\over 4})\delta^{\alpha,0}\delta^{\beta,0}\delta_{r+s,0}, \cr
[v^i_m, G^\alpha_r] &= \big[(\alpha+\half )m-(i+1)r\big]G^{\alpha+i}_{m+r},
\cr
[v^i_m, {\bar G}^\alpha_r] &=
\big[(\alpha+\half )m-(i+1)r\big]{\bar G}^{\alpha+i}_{m+r}, \cr
[J^{i-1}_m, G^\alpha_r] &=G^{i+\alpha}_{m+r}, \cr
     [J^{i-1}_m, {\bar G}^{\alpha}_r] &=-{\bar G}^{\alpha+i}_{m+r}.\cr
    [J^{i-1}_m, J^{j-1}_n] &= {c\over 2} m
                     \delta^{i,0}\delta^{j,0}\delta_{m+n,0}\cr }\eqno(42)
$$
 In fact, this is the algebra of symplectic diffeomorphisms on a
$(2,2)$ superplane, i.e. a plane of two bosonic and two fermionic dimensions
[30]. The $N=1$ super $w_\infty$ algebra can be obtained from the above
algebra by truncation, or directly as an algebra of the symplectic
diffeomorphisms of a $(2,1)$ superplane [30,31]. For $i=j=\alpha=\beta=0$, the
algebra (42) reduces to the well known $N=2$ superconformal algebra.

\bigskip
\centerline{\bf 7. $w_\infty$ \ Gravity}
\bigskip
It is natural to look for Lagrangian field theories in which the $W$
symmetries of the kind we have discussed so far are realised. In the
case of Virasoro symmetry, the classic example is the
Polyakov type Lagrangian
$$
{\cal L} =-{1\over 4}{\sqrt -h}h^{ij}\partial_i\phi
\partial_j\phi,   \eqno(43)
$$
where $h^{ij}$ is the inverse of the worldsheet metric
$h_{ij},\ (i,j=0,1),\ h=det h_{ij}$ and $\phi$ is a real scalar. This
Lagrangian clearly possesses the 2D diffeomorphism and Weyl symmetries.
In a conformal gauge the residual symmetry becomes the Virasoro symmetry.
Generalizations of this Lagrangian in which $W$ symmetry is realised, have been
constructed, and although they are not purely gauge theories of $W$ algebras,
they have been called $W$ gravity Lagrangians, motivated by the fact that they
are invariant under local $W$ symmetries. Here, we shall present selected few
examples of these  Lagrangian field  theories.

To generalize (43), it is convenient to first fix its Weyl symmetry
so that we are left with diffeomorphism symmetry. A convenient gauge choice
in which $h$ is a perfect square is as follows
$$
    h_{+-}={1\over 2}(1+h_{++}h_{--}),  \eqno(44)
$$
where the light-cone coordinate bases defined by $x^{\pm}=(x^0\pm x^1)$ is
used. In this gauge, the Lagrangian (43) reduces to
$$
{\cal L}={1\over 2}(1-h_{++}h_{--})^{-1}(\partial_+\phi-h_{++}\partial_-\phi)
(\partial_-\phi-h_{--}\partial_+\phi). \eqno(45)
$$
It turns out to be very useful to rewrite this Lagrangian in the following
first order form
$$
{\cal L}=-{1\over 2}\partial_+\phi \partial_-\phi -J_+J_- +J_+\partial_-\phi
+J_-\partial_+\phi-{1\over 2}h_{--}J_+^2-{1\over 2}h_{++}J_-^2, \eqno(46)
$$
where $J_{\pm}$ are auxiliary fields which obey the field equations
$$
     \eqalign{
              J_+ &=\partial_+\phi-h_{++}J_-, \cr
              J_- &=\partial_-\phi-h_{--}J_+, \cr}\eqno(47)
$$
These equations define a set of nested covariant derivatives [32]. Solving for
$J_{\pm}$ and substituting into (46) indeed yields (45). Thus the two
Lagrangians are classically equivalent, though in principal they may be quantum
inequivalent. The action of the Lagrangian (46) is invariant under 2D
diffeomorphism transformations, with a general parameter $k_+(x^+,x^-)$, given
by [13]  $$
\eqalign{
    \delta \phi &=k_+J_-  \cr
    \delta h_{++}&=\partial_+k_+-h_{++}\partial_-k_++k_+\partial_-h_{++} \cr
     \delta h_{--} &=0  \cr
     \delta J_- &=\partial_-(k_+J_-)   \cr
       \delta J_+ &=0,  \cr}\eqno(48)
$$
and 2D diffeomorphisms with parameters $k_-(x^+,x^-)$, which can be
obtained from the above transformations by changing  $+\leftrightarrow -$
everywhere.

The $W$ symmetric generalization of (46) is now remarkably simple. With the
further generalization to the case in which the fields $\phi$ and $J_{\pm}$
take their values in the Lie algebra of $SU(N)$ the answer can be written as
follows [13]
$$
  \eqalign{
           {\cal L}=& {\rm tr}\big(-{1\over 2}\partial_+\phi
\partial_-\phi-J_+J_- + J_+\partial_-\phi + J_-\partial_+\phi\big) \cr
   &-\sum_{\ell\ge 0}{1\over {\ell+2}}A_{+\ell}{\rm tr}J_-^{\ell+2}
 -\sum_{\ell\ge 0}{1\over {\ell+2}}A_{-\ell}{\rm tr}J_+^{\ell+2}.\cr}\eqno(49)
$$
 Note that $A_{0+}=h_{++}$ and $A_{0-}=h_{--}$. The equations of motion for the
auxiliary fields now reads
$$
     \eqalign{
              J_+ &=\partial_+\phi-\sum_{\ell\ge 0}A_{+\ell}J_-^{\ell+1}, \cr
       J_- &=\partial_-\phi-\sum_{\ell\ge 0}A_{-\ell}J_+^{\ell+1}.
\cr}\eqno(50)
$$
The Lagrangian (49) possesses the W-symmetry with parameters
$k_{+\ell}(x^+,x^-)$ that generalize (48) as follows
$$
\eqalign{
\delta\phi&=\sum_{\ell\ge -1}k_{+\ell}J_-^{\ell+1}\cr
\delta A_{+\ell}&=\partial_+ k_{+\ell} -
\sum_{j=0}^\ell [(j+1)A_{+ j}\partial_- k_{+(\ell-j)} -
(\ell-j+1)k_{+(\ell-j)} \partial_- A_{+ j}]\cr
\delta A_{-\ell}&=0  \cr
\delta J_-&=\sum_{\ell \ge -1}\partial_-\big[k_{+\ell}(J_-)^{\ell+1}\big]\cr
  \delta J_+&=0,  \cr}\eqno(51)
$$
and $W$ transformations with parameters $k_{+\ell}(x^+,x^-)$ which can be
obtained from above by the replacement $+\leftrightarrow -$ everywhere.
It is important to note that we must set
$$
    k_{-1\pm}=-{1\over N}\sum_{\ell\ge 1} k_{\pm\ell}{\rm
tr}(J_{\mp})^{\ell+1}, \eqno(52)
$$
to ensure the tracelessness $\delta \phi$ and $\delta J_{\pm}$ in the
transformation rules above.

The Lagrangian (49) has also Stueckelberg type shift  symmetries which
arise due to the fact that for $SU(N)$, only $(N-1)$ Casimirs of the form
${\rm tr}(J_{\pm})^{\ell+2}$ are really independent, while the rest can be
factorise into products of these Casimirs. There is not a simple way to write
down the resulting symmetries for any value of $N$; one must consider
each case separately. To see how this works, consider for example, the case of
$SU(3)$. It is sufficient to consider, say, the left-handed fields, since all
results hold independently also for the right-handed fields. We observe that,
for $SU(3)$,  the terms contating the left-handed gauge fields in the
Lagrangian
can be written as ( dropping the handedness subscripts)
$$
{\cal L}(A)=-{1\over 2}A_0{\rm tr}J^2 -{1\over 3}A_1{\rm tr}J^3
-\sum_{\ell\ge 2}{1\over \ell+2 }A_\ell X_\ell{\rm tr}J^2  + \sum_{\ell\ge
3}{1\over \ell+2} (A_\ell Y_\ell  {\rm tr} J^3 , \eqno(53)
$$
where $X$ and $Y$ are themselves pruducts of ${\rm tr}J^2$ and/or
${\rm tr}J^3$ factors. For example, since
${\rm tr} J^4={1\over 2}({\rm tr}J^2)^2$, this implies that
$X_2={1\over 2} {\rm tr}J^2$. It is now easy to see that the Lagrangian (49)
has
the following symmetry
$$
  \eqalign{
    \delta A_0 &= -\sum_{\ell\ge 2}{2\over {\ell+2}}\alpha_\ell X_\ell, \cr
    \delta A_1 &= -\sum_{\ell\ge 3}{3\over {\ell+2}}\alpha_\ell Y_\ell, \cr
    \delta A_\ell &= \alpha_\ell, \ \ell\ge 2,  \cr}
\eqno(54)
$$
where $\alpha_\ell(x^+,x^-)$ is an arbitrary parameter. This symmetry can be
used to set $A_{\pm\ell}=0$ for $\ell\ge 2$.  If we also set $k_{\pm\ell}=0$
for  $\ell\ge2$ then from (51) we see that $A_{\pm\ell},\ \ell\ge 3$ remain
zero  for all $k_{\pm0}$ and $k_{\pm1}$ transformations, while $A_{\pm2}$
remains  zero provided we make a compensating $\alpha_{\pm2}$ transformation
given by

$$
    \alpha_{\pm2}=2A_{\pm1}\partial_\mp k_{\pm1}-2k_{\pm1}\partial_\mp
A_{\pm1}. \eqno(55)
$$
This means that, while the transformation rule
for $A_{\pm1}$ receives no modifications, the transformation rule
for $A_{\pm0}$ does receive a modification given by
$$
(\delta  A_{\pm0})_{\rm extra}=-{1\over 2}\alpha_2 X_2, \eqno(56)
$$
where, as was noted above, we have $X_2={1\over 2} {\rm tr}J^2$. Putting all
these together one finds the gauge theory of nonchiral $W_3$, which can be
found in ref.~[13], which appears to be equivalent to that of ref.~[32]. (In
the
latter reference, the Lagrangian exhibits the
spin-2 symmetry in a covariant fashion, whereas in our case all spin
symmetries, including spin-2, are treated on equal footing.) Our Lagrangian
has additional symmetries, called the $\beta$ and $\gamma$ symmetries, whose
origin and role is not quite clear so far.

There exists an interesting chiral truncation of the $W$ gravity theory
discussed above. It is achieved by setting $A_{+\ell}=0$. In that case from
(50) we have $J_+=\partial_+\phi$ and $J_-=\partial_-\phi
-\sum_{\ell\ge 0}A_{-\ell}{\rm tr}(\partial_+\phi)^{\ell+1}$. In this case, it
is more convenient to work in second order formalism. Thus, substituting for
$J_{\pm}$ into the Lagrangian (49), we obtain [13]
$$
  {\cal L}= {1\over 2}{\rm tr}\partial_+\phi
\partial_-\phi -\sum_{\ell\ge 0}{1\over {\ell+2}}
A_{\ell}{\rm tr}(\partial_+\phi)^{\ell+2}, \eqno(57)
$$
where we have used the notation $A_{-\ell}=A_\ell$. This  Lagrangian has the
following  symmetry [13]
$$
\eqalign{
\delta\phi&=\sum_{\ell\ge -1}k_\ell(\partial_+\phi)^{\ell+1}\cr
\delta A_\ell&=\partial_- k_\ell -
\sum_{j=0}^{\ell+1} [(j+1)A_j\partial_+ k_{\ell-j}-
(\ell-j+1)k_{\ell-j} \partial_-+ A_j]\cr}\eqno(58)
$$
The Lagrangian (57) has also the appropriate Stueckelberg symmetry. Using
this symmetry one can obtain [13] the chiral $W_3$ gravity of ref.~[33]. Note
that the interaction term in this Lagrangian has the form of a  gauge field
$\times$  conserved current ${1\over {\ell+2}}{\rm
tr}(\partial_+\phi)^{\ell+2}.$ It is important to note that the OPE of these
currents do not form a closed algebra, while they do close with respect to
Poisson bracket. Hence, the $w_\infty$ symmetry described above is a
classical symmetry, as expected.

The nonchiral $w_\infty$ gravity decribed in (49), (50) and (51) is a
light-cone formulation of a covariant $w_\infty$ gravity which needs to be
defined. One way to do this is to introduce extra degrees of
freedom at each spin level, removable by newly introduced symmetries, so that
the gauge fields transform as irreducible representations of the 2D Lorentz
group . Thus, spin j gauge field can be represented by $e_\mu^{a_1...a_{j-1}}$
which is symmetric and traceless in the indices $a_1...a_{j-1}$. In the
light-cone basis, this means that they are all +'s or -'s. Since, the index
$\mu$ takes the values $\pm$, the gauge fields contain four components
at each spin level. Therefore, passage from the nonchiral theory
where the gauge fields have two components to a covariant theory, one needs to
introduce two extra components at each spin level. They can be removed by the
Lorentz and Weyl type symmetries present at each spin level. The action and
transformation rules of  covariant $w_\infty$ gravity obtained in this way can
be found in ref.~[34]. The same results were already derived in ref.~[35]
by gauging of the $w_\infty$ algebra.

It should be noted that $w_\infty$ symmetry is realised nonlinearly on the
scalar field $\phi$ in $w_\infty$ gravity theories described above. This
situation can be descibed in terms of a coset construction which provides a
geometrical picture. We have already noted that $w_{1+\infty}$ algebra can be
linearly realised as the algebra of symplectic diffeomorphisms of a cylinder.
Let the coordinates of the cylinder be $(x\equiv x^+, y)$. From (39) we see
that $y$ independent functions correspond to the spin-1 generator $v^{-1}_m$,
which is the only generator to be added to $w_\infty$ to obtain $w_{1+\infty}$.
Thus, suppressing the irrelevant $x^-$ dependence, the $y$-independent scalar
field $\phi(x)$ can be considered as parametrising the coset
$w_{1+\infty}/w_\infty $. (For simplicity, we consider the case of a
single scalar. The general case follows straightforwardly.) For a general
noninfinitesimal symplectic transformation, the action on a scalar field
$f(x,y)$ is given by $$ f\rightarrow {\tilde f}=e^{Ad_\Lambda}
f=f+\{\Lambda,f\}+ {1\over 2!}\{\Lambda,\{\Lambda,f\}\}+..., \eqno(59)
$$
where $\Lambda(x,y)$ is the arbitrary parameter, and the Poisson bracket is
defined in $(x,y)$ space. The coset representative is obtained by setting
$\Lambda(x,y)=\phi(x)$. Considering the action of the group element with an
infinitesimal parameter $\lambda$, the change in $\phi$ is found by standard
methods to be $$
     \delta \phi =\big(e^{-Ad_\phi}\lambda\big)_{y=0}, \eqno(60)
$$
where setting $y=0$ amounts to restriction to the coset direction. For the
choice $\lambda=k_\ell y^{\ell+1}$, this transformation rules yields the
promised result: $\delta\phi=k_\ell(\partial_+\phi)^{\ell+1}$. In a similar
fashion, it can be shown that the transformation rule of the scalar in the
nonchiral theory can be viewed as a special symplectic diffeomorphism of a four
dimensional manifold with coordinates $x^+,x^-,y,{\tilde y}$ [13].

{}From the commutation rules (37), it can be seen that the coset space
$w_{1+\infty}/ w_\infty $ is not a symmetric space. In particular,
the coset generators do not form a representation of the subalgebra
$w_\infty$. Hence, some features of the general theory of nonlinear
realisations are going to be non-standard. Instead, one could start from the
symmetric coset space $w_{1+\infty}^+ / Vir^+ $ [36], where the + on
$w_{1+\infty}$ indicates the restriction of the Fourier modes of the
generators $v^\ell_m$ to $m\ge -\ell-1$, while the + on Vir denotes
the restriction of the Virasoro generators $v^0_m$ to $m\ge -1$.
Here, the coset generators do form a representation of the subalgebra
$Vir^+$, but the coset is infinite dimensional, which means the intruduction of
infinitely many Goldstone fields. It turns out that, all but those which lie
at the ``left edge of the wedge'', i.e. $\phi^\ell_{-\ell-1}$ can be
eliminated by the imposition of suitable constraints [36]. In particular, the
transformation of the scalar $\phi^{-1}_0$ turns out to be exactly the
nonlinear transformation rule of the scalar $\phi$ we encountered above in
$w_\infty$ gravity [36]. However, now there is an infinite chain of scalar
fields, all of which are arising in a geometrical fashion, and this may open up
interesting new possibilities with regard to the field theoretic realisation
of $w_\infty$ symmetry.

        So far we have described the classical $w_\infty$ gravity. Recently,
quantum $w_\infty$ gravity was studied, and it was found that infinitely many
counterms are needed in order to remove matter dependent anomalies [37].
Remarkably, these counterterms correspond precisely to a renormalisation of
the classical $w_\infty$ currents to quantum $W_\infty$ currents. For
example, the classical spin-4 current ${1\over 4}(\partial\phi)^4$
renormalises to the quantum spin-4 current of the form given in (25). In
ref.~[37], it is further shown that the matter independent gauge anomalies are
cancelled by the ghost contributions.

\bigskip
\centerline{\bf 8. $W_\infty$ \ Gravity}
\bigskip
The  gauge field $\times$  conserved current form of the chiral $w_\infty$
gravity Lagrangian discussed above suggests a $W_\infty$ gravity Lagrangian
of similar form. To this end, let us consider the following Lagrangian,
$$
 \lagr= \partial_+ \f^* \partial_- \f +\sum_{i\ge 0} A_i V^i, \eqno(61)
$$
where $V^i$ now represent the $W_\infty$ current given in (7). This Lagrangian
is indeed $W_\infty$ symmetric. The transformation rule for the scalar can be
derived by considering its  Poisson bracket with the current. Equivalently,
when appropriate (see below), one can instead use the OPE rules. For example,
using OPE, the transformation rule for the scalar field can be  obtained as
follows  $$
\eqalign{
           \delta_\phi &=\oint{dz\over 2\pi i}k_i(z)V^i(z) \phi  \cr
                 &=\sum_{i\ge 0}\sum_{k=0}^i \ \alpha^i_k\partial_+^k(k_i
\partial_+^{i+1-k} \phi),   \cr}\eqno(62)
$$
where the basic OPE rule (3) has been used. To derive the transformation
property of the gauge fields, it is convenient to first derive the
transformation of the curent $V^i$ and then demand the invariance of the
Lagrangian (61). To this end we need to know the OPE of two currents. This is
known to be [1,10,14]
$$
V^i(z) V^j(w)\sim -\sum_{\ell=0}^\infty {f^{ij}_{2\ell}}
(\partial_z,\partial_w)
\left({V^{i+j-2\ell}(w)\over z-w}\right)+{\rm central\ terms}, \eqno(63)
$$
where
$$
\eqalign{
           f_{\ell}^{ij}(m,n) &={1\over
2(\ell+1)!}\,\phi_{\ell}^{ij}\, M_{\ell}^{ij}(m,n), \cr
M^{ij}_{\ell}(m,n) &\equiv\sum_{k=0}^{\ell+1}(-1)^k
{\ell+1\choose k}
     (2i-\ell+2)_k[2j+2-k]_{\ell+1-k}\,m^{\ell+1-k}n^k.\cr}\eqno(64)
$$
Note that $M^{ij}_{\ell}(m,n)$ is obtained from $N^{ij}_{\ell}(m,n)$ by
picking up the leading component, with total  degree $(\ell+1)$ in $m$ and $n$.

To derive the {\it classical} tranformation rule of the current
$V^i$ which is bilinear in scalar field, we must consider only the singular and
field dependent terms in the OPE above, since that would correspond
to a single contraction of the scalar field, giving results equivalent to those
obtainable by using the Poisson bracket. (The central terms which we need to
omit  correspond to double contraction of the scalar fields.) With this in
mind, we obtain
$$
\eqalign{
            \delta V^i &=\oint{dz\over 2\pi i}k_i(z)V^i(z) V^j(\omega) \cr
       &=\sum_{\ell\ge 0}  \int dz\, k_j(z)\,
f^{ij}_{2\ell}(\partial_\omega, \partial_z)\Big(\delta(z-\omega)
V^{i+j-2\ell}(\omega)\Big). \cr} \eqno(65)
$$
The invariance of the Lagrangian (61) is now seen to be achieved by
letting the gauge fields to transform as
$$
\delta A_i=\partial_-k_i + \sum_{\ell\ge0}\sum_{j=0}^{i+2\ell}
f^{j, i-j+2\ell}_{2\ell}(\partial_A,
\partial_k) A_j k_{i-j+2\ell},  \eqno(66)
$$
where $\partial_A$ and $\partial_k$, which replace $m$ and $n$, are the
$\partial_+$ derivatives acting on $A$ and $k$, respectively.  Writing
$\delta A_i=\partial_- k_i +\hat\delta A_i$, we observe that
$\hat \delta A_i$ is a co-adjoint transformation of $A_i$, while $V^i$
transforms in the adjoint representation of $W_\infty$; i.e.
$\int\Big(\hat\delta A_i V^i+ A_i \delta V^i\Big)=0$.  Unlike the case of
finite-dimensional semi-simple Lie algebras, one must distinguish between the
adjoint and co-adjoint representations of $W_\infty$ since there is no
Cartan-Killing metric available to raise or lower indices.   One consequence of
this is that whereas the transformation (65) of $V^i$ under $k_j$ for given $i$
and $j$ involves a finite number of terms, the transformation (66) of $A_i$
under $k_j$ will involve  an infinite number.
      The simultaneous gauging of both the left-handed and the right-handed
$W_\infty$ symmetries is also possible. The results, which are rather
analogous to those of $w_\infty$ gravity, can be found in [14].

\bigskip
\centerline{\bf 9. Various Applications of $w_\infty$ and $W_\infty$ Algebras}
\bigskip
So far we have primarily discussed the structure of various $W_\infty$
algebras,
and their gauging. There are various applications of $w_\infty$ or $W_\infty$
algebras where these structures are relevant. The ultimate application would
be the construction of a W-string or W-membrane theory. Below we shall
comment briefly on a selected few topics with regard to the application of
W-symmetry.

\noindent 1. $w_\infty$ arises as a symmetry algebra of
$sl(\infty)$ Toda theory, which itself arises from a particular reduction of
the 4D self-dual gravity equation [8]. In ref.~[8], these results are obtained
from a Wess-Zumino-Witten  model by Hamiltonian reduction procedure. It should
also be pointed out that self-dual metrics of a spacetime with signature
$(2,2)$ are the ones that arise as consistent backgrounds for $N=2$
supersymmetric string theory [9].

\noindent 2. It has been shown that the Poisson bracket algebra
of an important, integrable hierarchy of nonlinear partial differeantial
equations known as the KP hierarchy, is isomorphic to the $W_{1+\infty}$
algebra [24]. Furthermore, it is known that another $W_{1+\infty}$ algebra,
which commutes with the first one arises as the symmetry algebra of the KP
hierarchy [4,5]. These facts play an important role in the nonperturbative
analysis of $d=1$ string theory.

The KP hierarchy is neatly  formulated in terms of the pseudo-differential
operator
$$
    Q=\partial_x+q_0\partial^{-1}_x+q_1\partial^{-2}_x+q_2\partial^{-3}_x
+\cdots    \eqno(67)
$$
where $q_i,\ i=0,1,2,...$ are functions of $x_1\equiv x,\ x_2\equiv y,\
x_3\equiv t$ and the ``higher time variables'' $x_4,x_5,...$. Defining the
Hamiltonians as
$$
    {\cal H}_n ={1\over {n+1}}\int res\ Q^{n+1}, \eqno(68)
$$
where the residue is defined to be the coeffient of $\partial^{-1}$, the KP
equations are simply the Hamiltonian flows
$$
    {\partial q_i\over {\partial x_n}}=\{q_i,\ {\cal H}_n\}.\eqno(69)
$$
with respect to the following Poisson bracket
$$
    \{q_i(x),\ q_j(x')\}=K_{ij}\delta(x,x')  \eqno(70)
$$
where [38]
$$
   K_{ij}=\sum_{\ell=0}^{i}(-1)^\ell{i\choose \ell}q_{i+j-\ell}\partial_x^\ell
            -\sum_{\ell=0}^j{j\choose \ell}\partial^\ell_x q_{i+j-\ell}.
       \eqno(71)
$$
The $n=1$ equation is content-free, while the $n=2,3$ equations together imply
the well known KP equation $ \partial_x(u_t+6uu_x+u_{xxx})=\pm 3u_{yy}$ where
$u\equiv q_0,\ u_t\equiv {\partial\over \partial t},\ u_x\equiv
{\partial\over \partial x}$, etc.  The flows (69) commute, i.e.
$\{{\cal H}_m,\ {\cal H}_{n}\}=0$, which shows that the KP hierarchy is
integrable. It has been shown that the Poisson bracket algebra (70) is
isomorphic to the OPE algebra of fermionic bilinears $:{\bar
\psi}\partial^{i-1}\psi:$, i.e. the following correspondence  exists [24]
$$
  \{q_i(x),\ q_j(x')\}\leftrightarrow [:{\bar
\psi}\partial^{i-1}\psi:,\ :{\bar \psi}\partial^{j-1}\psi:]  \eqno(72)
$$
On the other hand, the OPE algebra of these bilinears is just the
$W_{1+\infty}$ algebra. There exists another $W_{1+\infty}$ symmetry which
commutes with this one, and arises as the non-isospectral flow of the KP
potentials $q_i$ [4,5]. We refer the reader to [5] for a discussion of how this
symmetry can be utilized in the formulation of the matrix
model approach to 2D gravity. A basic connection is the identification of the
partition function of the 2D gravity with a suitable function of certain KP
potentials.

\noindent 3. Perhaps one of the most important applications of $W$ symmetry is
the construction of a possible $W$ string theory. So far, most of the
studies on $W$ algebras have primarily aimed at a better understanding of their
algebraic structure, and there have been very few attempts to build a $W$
string theory. This, indeed, seems to be a rather nontrivial challenge. In [3],
various aspects of a possible $W$ string theory were conjectured, with
emphasis on $W_3$. The critical dimension for $W_3$ string was suggested to
be $d=100$. In particular, it was conjectured that {\it massless} target space
higher spin fields should occur in the spectrum. More recently, the issues of
physical states in a $W_N$ string theory with emphasis on $W_3$ has been
investigated withing the famework of a specific matter coupling to $W$ gravity
[39].  Consider $d$ free scalars $X^i$, and the fields $\phi_1,\phi_2$ coming
from the $W_3$ gravity sector. The spin-2 generator of $W_3$ algebra is the
total energy momentum tensor given by
$$
\eqalign{
      T(z) &=-{1\over 4}(\partial\phi_1)^2
+ia_1\partial^2\phi_1-{1\over 4}(\partial\phi_2)^2+ia_2\partial^2\phi_2-
{1\over 2}(\partial{\vec X})^2 \cr
      &\equiv T_1+T_2+T_X  \cr}\eqno(73)
$$
where $a_1^2=(12\alpha_0^2+c_m)/24$, $c_m$ is the central charge of the matter
fields (i.e. $d$, for $d$ free scalars), $\ a_2^2=3\alpha_0/2$ and
$\alpha_0=-49/4$ [39]. The observation made in ref.~[39] is that, since
$\phi_1$
occurs only through its energy-momentum tensor $T_1$ in the expression for the
spin-3 current of pure $W_3$ gravity, a natural way to couple matter fields
$X^i$ is to make the replacement $T_1\rightarrow T_1+T_X$ in that expression.
In this way one arrives at the following spin-3 generator [39]
$$
W={b\over 12i}\big(
(\partial\phi_2)^3-6ia_2(\partial\phi_2)(\partial^2\phi_2)+
12\partial\phi_2(T_1+T_X)-12ia_2\partial(T_1+T_X), \eqno(74)
$$
where $b^2={16\over {22+10(1-24\alpha_0^2)}}$. Ref.~[39] then considers
the following operator which creates a {\it scalar} (tachyon) mode of the
string
$$
 V_T(z)=e^{i{\vec \beta}\cdot{\vec \phi}}e^{i{\vec k}\cdot{\vec X}}.\eqno(75)
$$
Phsical state conditions imposes restrictions on ${\vec \beta}$ and
${\vec k}$. It turns out that a tachyon occurs for $d>{1\over 2}$, and thus
the lower critical dimension is $d={1\over 2}$ [39]. The upper critical
dimension in which a full symmetry betwen the usual coordinates and the $W_3$
gravity coordinate arises turns out to be $d={49\over 2}$ [39]. The full
spectrum and other salient features of an off-critical  $W$ string is not known
at present.

Then, there is the question of what an $W_\infty$ string theory would
be like. Those fetaures which arise in $W_N$ string  as a consequence of the
nonlinear nature of the $W_N$ algebra may disappear, and new features having to
do with the infinite $N$ limit may arise. We would like to conjecture that
in a super $W_\infty$ string theory, the target space spectrum will contain an
infinite tower of higher spin gauge fields, gauging an infinite dimensional
higher spin algebra which contains the usual Poincar\'e or anti
de Sitter superalgebras as a subalgebra. If we assume that the motivations of
Fradkin and Vasiliev [40] for considering the anti de Sitter rather then
Poincar\'e based higher spin algebras persist also in a $W_\infty$ string
theory, then it is reasonable to try  construct the generators of the
Fradkin-Vasiliev type higher spin algebra from the $W_\infty$
structures on the worldsheet, and show that they obey the correct algebra at
the quantum level.

\noindent 4. Symplectic diffeomorphisms of a surface also arises in membrane
theories as a residual symmetry group in a light-cone gauge [41]. Consider the
membrane action
$$
   I=\int d\tau d\sigma d\rho \bigl ( {1\over 2}{\sqrt -g} g^{ij}
   \partial_i X^\mu\partial_j X^\nu \eta_{\mu\nu} - {1\over 2}{\sqrt -g}\bigr
),
                                                                     \eqno(76)
$$
where $i=\tau,\sigma,\rho$ labels the coordinates of the membrane world-volume
with metric $g_{ij}$, and $X^\mu$, where $\mu =0,...,d-1$, are the coordinates
o
   f a
d-dimensional Minkowski spacetime with metric $\eta_{\mu\nu}$. The action is
evidently invariant under the reparametrization of the world-volume, $\delta
\sigma^i = \xi^i$. Imposing the following gauge conditions [42]
$$
  X^+ =\tau, \ \ \ \ g^{00}=-h^{-1},  \eqno(77)
$$
where $X^+ = {1\over \sqrt 2}(X^0 + X^{d-1}),\ h=det g_{ab}$, and
$a=\sigma,\rho $, we are left as a residual symmetry with the symplectic
diffeomorphisms of the membrane satisfying the condition  $\partial_a \xi^a
=0$. To maintain the above gauges, one finds that, at least locally,
$g^{0a}$ must have the form
$$
g^{0a}=-\epsilon^{ab} h^{-1} \partial_b \omega,  \eqno(78)
$$
where $\omega$ is a time-dependent gauge potential which transforms as
$$
 \eqalign{\delta \omega &= \partial_0 \Lambda +
           \epsilon^{ab}\partial_b \omega \partial_a \Lambda  \cr
  &\equiv \partial_0 \Lambda +\{\omega, \Lambda\} \equiv D_0 \Lambda.
    \cr} \eqno(79)
$$
Here  $\Lambda(\tau,\sigma,\rho)$ is an arbitrary gauge parameter. With the
gauge choices (77), and recalling (78), one can show that the membrane action
(76) reduces to
$$
      I=\int d\tau \int d\sigma d\rho \bigl[{1\over 2} D_0 X^r D_0 X^r
   -{1\over 4}\{X^r, X^s\}\{X^r, X^s\} \bigr], \ \ \ r,s=1,...,d-2,  \eqno(80)
$$
where $ D_0 X^r = \partial_0 X^r +\{\omega, X^r\}$. The action (76) is
invariant
under
$$
    \delta \omega = D_0 \Lambda, \ \ \ \ \ \delta X^r = -\{\Lambda , X^r \}.
    \eqno(81)
$$
Thus we see that the
action (76) is the gauge theory in one dimension (time) of the
symplectic diffeomorphisms of the membrane. The role of the usual trace is
played  by $\int d\sigma d\rho$, and Yang-Mills commutator is replaced by the
Poisson bracket $\{,\}$. Typically spherical or toroidal membranes have been
considered so far. Perhaps, one should consider a cylindirical membrane, and
investigate the role of $w_\infty$ symmetry in the theory. It would be
interesting to see if the membrane theory is integrable under any circumstances
and if the representation theory of $w_\infty$ could be utilized in analysing
the quantum theory. It would also be interesting to determine the quantum fate
of $w_\infty$ symmetry and see whether it deforms to $W_\infty$ symmetry in a
manner similar to the one encountered in the study of $w_\infty$ gravity [37].

\bigskip\bigskip
\centerline{\bf Acknowledments}
\bigskip\bigskip
I would like to thank E. Bergshoeff, A. Das, C. Pope, X. Shen, and S. Sin
for useful discussions. I also would like to thank Professor Abdus Salam, the
International Atomic Energy Agency and UNESCO for hospitality at the
International Center for Theoretical Physics where this work was completed.
This work is partially supported by NSF grant PHY-8907887.

\vfill\eject

\bigskip

\centerline{\bf References}
\bigskip

\frenchspacing

\item{1.} C.N. Pope, L.J. Romans and X. Shen, Phys. Lett. {\bf 236B} (1990)
173.
\item{} C.N. Pope, L.J. Romans and X. Shen, Nucl. Phys. {\bf B339} (1990) 191.
\item{2.}E. Brezin and V.A. Kazakov, Phys. Lett. {\bf 236B} (1990) 144;
\item{} D.J. Gross and A.A. Migdal, Phys. Rev. Lett. {\bf 64} (1990) 127;
\item{} M. Douglas and S. Shenker, Nucl. Phys. {\bf B335} (1990) 635.
\item{} M. Douglas, Phys. Lett. {\bf 238B} (1990) 176.
\item{3.} A. Bilal and J.L. Gervais, Nucl. Phys. {\bf B326} (1989) 222.
\item{4.}A.Yu. Orlov, in {\it Plasma theory and nonlinear and turbulent
processes in physics, Vol. 1}, Eds. V.G. Bar'yakhtar et al (World Scientific,
1988).
\item{5.} M.A. Awada and S.J. Sin, {\it Twisted $W_\infty$ symmetry and
the KP hierarchy and the string equation of d=1 matrix models}, preprint,
UFIFT-HEP-90-33 (November 1990).
\item{6.}I. Bakas, Phys. Lett. {\bf 228B} (1989) 57.
\item{7.}  A. Bilal, Phys. Lett. {\bf 227B} (1989) 406.
\item{8.} Q.H. Park, Nucl. Phys. {\bf B333} (1990) 267; {\it 2D sigma model
approach to 4D instantons}, preprint, UMDEPP 90-270 (September 1990).
\item{9.} H. Ooguri and C. Vafa, Mod. Phys. Lett. {\bf A5} (1990) 1389; {\it
Geometry of N=2 strings}, preprint, EF1 91/05, HUTP-91/A003 (January 1991).
\item{10.}C.N. Pope, L.J. Romans and X. Shen, Phys. Lett. {\bf 242B} (1990)
401.
\item{11.} E. Bergshoeff, C.N. Pope, L.J. Romans, E. Sezgin and X. Shen,
Phys. Lett. {\bf 245B} (1990) 447.
\item{12.} C.N. Pope, L.J. Romans, E. Sezgin and X. Shen, Phys. Lett.
{\bf 256B} (1991) 191.
\item{13.} E. Bergshoeff, C.N. Pope, L.J. Romans, E.
Sezgin, X. Shen and  K.S. Stelle, Phys. Lett. {\bf 243B} (1990) 350.
\item{14.} E. Bergshoeff, C.N. Pope, L.J. Romans, E. Sezgin and X. Shen, Mod.
Phys. Lett. {\bf A5} (1990) 1957.
\item{15.} P. Ginsparg, {\it Applied conformal field theory}, Les Houches
lectures, 1988.
\item{16.} A.B. Zamolodchikov, Teo. Mat. Fiz. {\bf 65} (1985)
347. \item{17.} V.A. Fateev and S. Lykyanov, Int. J. Mod. Phys. {\bf A3} (1988)
507. \item{18.}  I. Bakas and E. Kiritsis, Nucl. Phys. {\bf B343} (1990) 185.
\item{19.}  I. Bakas and E. Kiritsis, in {\it Topological Methods in Quantum
Field Theory}, Eds. W. Nahm, S. Randjbar-Daemi, E. Sezgin and E. Witten
(World Scientific, 1991).
\item{20.} C.N. Pope, L.J. Romans and X. Shen, Phys. Lett. {\bf 245B} (1990)
72.
\item{21.} E. Witten, private communication; Comm. Math. Phys. {\bf 113}
(1988) 529.
\item{22.} A. Zamolodchikov, private communication.
\item{23.} M. Fukuma, H. Kawai and R. Nakayama, {\it Infinite dimensional
grassmannian structure of two dimensional gravity}, preprint, UT-572,
KEK-TH-272 (November 1990).
\item{24.} K. Yamagishi, Phys. Lett. {\bf 259B} (1991) 436.
\item{25.} E. Bergshoeff, B. de Wit and M. Vasiliev, {\it The structure of
the super $W_\infty(\lambda)$ algebra}, preprint, CERN-TH.6021/91, THU-91/05
(April 1991).
\item{26.} E.S. Fradkin and V. Ya.Linetsky, Mod. Phys. Lett. {\bf A5}
(1990)1167; J. Math. Phys. {\bf 32} (1991) 1218; {\it Supersymmetric version of
3D Virasoro-like algebra and supersymmetrization of $SU(\infty)$}, preprint,
DSF-T-90-10 (September 1990).
\item{27.} S. Odake and T.
Sano, {\it $W_{1+\infty}$ and super $W_\infty$ algebras with $SU(N)$ symmetry},
preprint, UT-569 (October 1990).
\item{28.} I. Bakas and E. Kiritsis, in {\it
Common trends in mathematics and quantum field theories}, Eds. T. Eguchi, T.
Inami and T. Miwa (Prog. Theor. Phys. Supplement, 1990).
\item{29.} I. Bars, C.N. Pope and E. Sezgin, Phys. Lett. {\bf 210B} (1988) 85.
\item{30.} C.N. Pope and X. Shen, Phys. Lett. {\bf 236B} (1990) 21.
\item{31.} E. Sezgin, in {\it Strings '89}, Eds. R. Arnowitt, R. Bryan,
M.J. Duff, D. Nanopoulos and C.N. Pope (World Scientific, 1990).
\item{32.} K. Schoutens, A. Sevrin and P. van Nieuwenhuizen, Phys. Lett. {\bf
243B} (1990) 245.
\item{33.} C.M. Hull, Phys. Lett. {\bf 240B} (1989) 110.
\item{34.} E. Bergshoeff, C.N. Pope and K.S. Stelle, Phys. Lett. {\bf 249B}
(1990) 208.
\item{35.} K. Schoutens, A. Sevrin and P. van Nieuwenhuizen, Phys. Lett.
{\bf 251B} (1990) 355.
\item{36.} E. Sezgin and K.S. Stelle, {\it Nonlinear realisations of
$w_{1+\infty}$}, preprint, CTP TAMU-26/91, Imperial/TP/90-91/21.
\item{37.} E. Bergshoeff, P.S. Howe, C.N. Pope, E. Sezgin, X. Shen
and K.S. Stelle, {\it Quantisation deforms $w_\infty$ to $W_\infty$ gravity},
preprint, CTP TAMU-25/91, Imperial/TP/90-91/20, ITP-SB-91-17 (April 1991).
\item{38.} Y. Watanabe, Lett. Math. Phys. {\bf 7} (1983) 99.
\item{39.}  S.R. Das, A. Dhar and S.K. Rama, {\it Physical states and scaling
properties of W gravities and W strings}, preprint, TIFR/TH/91-20 (March
1991).
\item{40.} E.S. Fradkin and M. Vasiliev, Ann. Phys. {\bf 177} (1987) 63.
\item{41.} J. Hoppe, MIT Ph. D. Thesis, 1982 and in {\it Proc. Int. Workshop
            on Constraints Theory and Relativistic Dynamics},
            Eds. G. Longhi and L. Lusanna (World Scientific, 1987).
\item {42.} E. Bergshoeff, E. Sezgin, Y. Tanii and P. Townsend, Ann. Phys.
            {\bf 199} (1990) 340.

\bye